\documentclass[journal]{IEEEtran}
\usepackage{graphicx}
\usepackage{amsmath}
\usepackage{subfigure}
\usepackage{rotating}
\usepackage{multirow}
\usepackage{array}
\usepackage{rotating}
\usepackage{auto-pst-pdf}
\usepackage{amsmath}

\begin{document}
\title{Modeling Propagation Characteristics for Arm-Motion in Wireless Body Area Sensor Networks}

\author{Q. Ain, A. Ikram, N. Javaid, U. Qasim$^{\ddag}$, Z. A. Khan$^{\S}$\\
        $^{\ddag}$University of Alberta, Alberta, Canada\\
        Department of Electrical Engineering, COMSATS\\ Institute of
        Information Technology, Islamabad, Pakistan. \\
        $^{\S}$Faculty of Engineering, Dalhousie University, Halifax, Canada.
        }

\maketitle

\begin{abstract}

To monitor health information using wireless sensors on body is a promising new application. Human body acts as a transmission channel in wearable wireless devices, so electromagnetic propagation modeling is well thought-out for transmission channel in Wireless Body Area Sensor Network (WBASN). In this paper we have presented the wave propagation in WBASN which is modeled as point source (Antenna), close to the arm of the human body. Four possible cases are presented, where transmitter and receiver are inside or outside of the body. Dyadic Green's function is specifically used to propose a channel model for arm motion of human body model. This function is expanded in terms of vector wave function and scattering superposition principle. This paper describes the analytical derivation of the spherical electric field distribution model and the simulation of those derivations.
\end{abstract}


\section{Introduction}
Hospitals throughout the world are facing a unique problem, as the aged population is increased, health-care population is decreased. Telecommunication community is not doing much work in the field of medicine however, there is a need of remote patient monitoring technology. To fulfill this task, it is required to build communication network between an external interface and portable sensor devices worn on and implemented within the body of the user which can be done by BASNs.

BASNs is not only useful for remote patient monitoring, but can also establishes within the hospitals; like in operation theaters and intensive care units. It would enhance patient comfort as well as provide ease to doctors and nurses to perform their work efficiently. BAN is used for connecting body to wireless devices and finds applications in various areas such as entertainment, defense forces and sports.

The basic step in building any wireless device is to study the transmission channel and to model it accurately. Channel modeling is a technique that has been initiated by a group of researchers throughout the world [1]. They have studied path loss and performed measurement campaigns for wireless node on the body [2-8]. Some researchers have taken into account, the implanted devices which are the area of BAN called as intra-body communication [9]. For the short range low data rate communication in BAN, measurement groups have considered Ultra-Wide Band (UWB) as the appropriate air interface. The models developed by measurement campaigns are only path loss models and do not provide any description of propagation channel.

It is important to study the propagation mechanism of radio waves on and inside the body in order to develop an accurate BAN channel model.  This study will show the underlying propagation characteristics. It would help in the development of BAN transceivers which are much suited to the body environment.

For a given position of the transmitter on or inside the body it is required to find out the electromagnetic field on or inside the body for a BAN channel model. This is quite a critical problem that requires a large amount of computational power. Therefore, it is necessary to derive an analytical expression which will perform this objective. In short this determines which propagation mechanism takes place, that is reflection, diffraction and transmission [10]. An appropriate method of doing this task is by using Dyadic Green's function. The solution of canonical problems, such as cylinder, multi layer and sphere have been solved in Electro Magnetic (EM) theory, using Dyadic Green's Functions [11-13].

\section{Motivation}

Recently, WBASNs shows potential due to increasing application in medical health care. In WBASNs, each sensor in the body sends it's data to antenna,both sensors and antenna are worn directly on the body. Examples include sensors which can measure Brain activity, blood pressure, body movement and automatic emergency calls. We require simple and generic body area propagation models to develop efficient and low power radio systems near the human body. To achieve better performance and reliability, wave propagation needs to be modeled correctly. Few studies have focused on analytic model of propagation around a cylinder (as human body resembles a cylinder) using different functions. These functions involve Mathieu function, Dyadic Green's function, Maxwell's equations, Finite Difference Time Domain (FDTD) and Uniform Theory of Diffraction (UTD). Some of these approaches have already proven effective for evaluating body area communication system proposals.

Finite Difference Time Domain had successfully measured the communication scenarios. Complete Ultra-Wide band models have been developed using measurements and simulations, however they do not consider the physical propagation mechanism. So, the researchers have to rely on ad-hoc modeling approaches which can result in less accurate propagation trends and inappropriate modeling choices [14, 15].

Uniform Theory of Diffraction depends on a ray tracing mechanism allowing propagation channel to be explained in terms of ray diffraction around the body . It typically based on high-frequency approximations which is not valid for low frequencies, also not useful when antenna is very close to the body [16].

A generic approach is proposed to understand the body area propagation by considering the body as a lossy cylinder and antenna as a point source by using Maxwell's equation. A solution for a line source near lossy cylinder is derived using addition theorem of Hankel functions then the line source is converted into the point source by taking inverse Fourier transform. The model accurately predicts the path loss model and can be extended to all frequencies and polarities but this is limited in scope and not always physically motivated [17].

Mathieu functions are also used for body area propagation model. The human body is treated as a lossy dielectric elliptic cylinder with infinite length and a small antenna is treated as three-dimensional (3-D) polarized point source. First the three-dimensional problem of cylinder is resolved into 2-D problem by using Fourier transform and then this can be expanded in terms of Eigen functions in cylindrical coordinates. By using Mathieu function exact expression of electric field distribution near the human body is deduced [18].

The propagation characteristics of cylindrical shaped human body have been derived using Dyadic Green's functions. The model includes the cases of transmitter and receiver presents either inside or outside of the body and also provides simulation plots of Electric field with different values of angle $(\theta)$.
All the above proposals describe the propagation characteristics of cylindrically shaped human model [19].

We have developed a simple but generic approach to body area propagation derived from Dyadic Green's Function (DGF). This approach is for arm motion of human body. When the human arm is moved in $ r,\theta,\phi$ direction, propagation characteristics of spherical shaped have been derived using DGF. First, we use spherical vector Eigen functions for finding the scattering superposition. Four cases are considered for either transmitter or receiver is located inside or outside the body. Finally, simulated results of electric field distribution with different values of angle have shown.

\section{Mathematical Modeling for Arm Motion using Dyadic Green's Function}

In this paper, spherical symmetry is used to represent in and around the arm of the human body. A point on body is a sensor, denoted by x which represents ($r$,$\Theta$,$\phi$) coordinates in the spherical coordinate system and $x_0$  is the location of transmitting antenna. ($r$,$\Theta$,$\phi$) are unit vectors along radial, angle of elevation from z-axis and azimuthal angle from x-axis as shown in figure 1.

\begin{figure}[t]
\centering
\includegraphics[height=7cm, width=8cm]{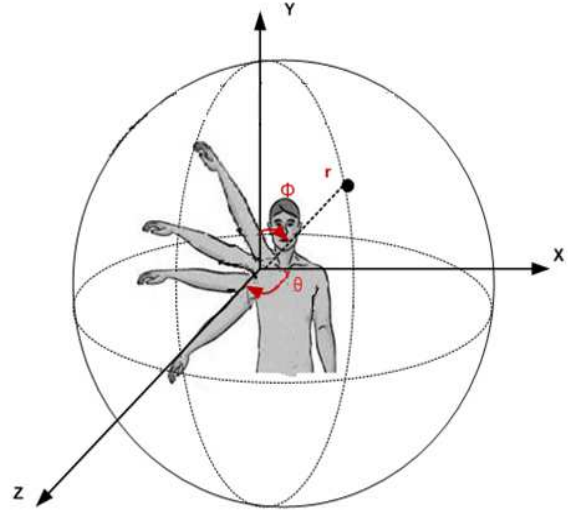}
\caption{ \textbf{Human body model showing arm motion in 3D.}}
\end{figure}

\subsection{Electric Field Propagation Characteristics}

Let $E(x)$ be electric field at point $x$ due to current source $J(x_0)$. The general formula for Electric field can be written as:

\begin{eqnarray}
E(x)=i \omega \mu_{p} \int \int \int_{V} G(x,x_{o})J(x,x_{0})dv
\end{eqnarray}

$V$ is volume of source, $J(X_0)$ is the current source, $G(x,x_0)$ is the Dyadic Green's function $'\omega'$ is the radian frequency of transmission and $'\mu_{p}'$ is magmatic permeability of the medium. A Dyadic Green's function is a type of function used to solve inhomogeneous differential equations subject to specific initial conditions or boundary condition.

\subsection{Spherical Wave Vector Eigen Function}

As we are considering arm motion of human body, so spherical symmetry is used by taking shoulder as center. For this, spherical eigen functions  are used to write the Dyadic Green's function.

Dyadic Green's function is basically depends on the spherical vector eigen functions [14]. These eigen functions are $L_{nhk}(\chi)$, $M_{nhk}(\chi)$ and $N_{nhk}(\chi)$, where $k$ is the wave number of medium, $n$ is an integer, $h$ is a real number and $x$ is a point in space. These all are the solutions to the Helmholtz equation having three components in $r$, $\Theta$ and $\phi$. These vector eigen functions are given by [19]:

\begin{eqnarray}
L_{nhk} (\chi)=\nabla[\Psi_{nhk}(\chi)]
\end{eqnarray}

\begin{eqnarray}
M_{nhk}(\chi)=\nabla\times[\Psi_{nhk}(\chi)]
\end{eqnarray}

\begin{eqnarray}
N_{nhk}(\chi)=\frac{1}{k}\nabla\times\nabla[\Psi_{nhk}(\chi)]
\end{eqnarray}

In above eigen functions, Laplacian operator in the spherical coordinate system is $\nabla$. It's mathematical expression is given as:

\begin{eqnarray}
\nabla=\frac{\partial}{\partial r} + \frac{\partial}{r \partial \theta} + \frac{\partial}{r \sin \theta  \partial \phi}
\end{eqnarray}

$x$ represents the point in space having components $r$, $\Theta$ and $\phi$. Solution of Helmoltz equation is $\Psi_{nhk}(x)$ which is the scalar eigen function [19].

\begin{eqnarray}
[\Psi_{nhk}(\chi)]= Z_{n}(\eta r) P^{h}_{n}(\cos\theta)_{\sin}^{\cos} h \phi
\end{eqnarray}

$Z_{n}$ is a general spherical function of order $n$. For sphere we use Hankle function of first and second order which are defined as:

\begin{eqnarray}
[Z_{n}(\eta r)]= (-1)^{n}(\eta r)(\frac{d}{dr\eta^{2} r})^n(\frac{\sin(\eta r)}{\eta r})^{n}
\end{eqnarray}

$\eta$ is the propagation constant in direction of $\phi$, whereas $k^2=\eta^2 + h^2$. The laplace operator is applied and find the eigen values $L_{nhk}$, $M_{nhk}$ and $N_{nhk}$ by using eigen function. The vector eigen function in (2), (3) and (4) becomes:

\begin{eqnarray}
\begin{split}
L_{nhk}(\chi)=\frac{\partial Z_{n}(\eta r)}{\partial r}P^{h}_{n}(\cos\theta)_{\sin}^{\cos} h \phi+
\frac{z_{n}(\eta r)}{r}\\\frac{\partial}{\partial\theta}P^{h}_{n}(\cos\theta)_{\sin}^{\cos} h \phi +
 \frac{h Z_{n}(\eta r)}{r\sin\theta}P^{h}_{n}(\cos\theta)_{\cos}^{\sin} h \phi
 \end{split}
\end{eqnarray}

\begin{eqnarray}
\begin{split}
M_{nhk}(\chi)=\mp\frac{h Z_{n}(\eta r)}{\sin\theta}P^{h}_{n}(\cos\theta)_{\cos}^{\sin} h \phi-
 Z_{n}(\eta r) \\\frac{\partial}{\partial\theta}P^{h}_{n}(\cos\theta)_{\sin}^{\cos} h \phi
\end{split}
\end{eqnarray}

\begin{eqnarray}
\begin{split}
N_{nhk}(\chi)=\frac{n Z_{n}(\eta r)}{k r}P^{h}_{n}(\cos\theta)_{\sin}^{\cos} h \phi +
\frac{1}{k r} \\\frac{\partial r Z_{n}(\eta r)}{\partial r}P^{h}_{n}(\cos\theta)_{\sin}^{\cos} h \phi
 \mp \frac{h}{\sin\theta}P^{h}_{n}(\cos\theta)_{\cos}^{\sin} h \phi
 \end{split}
\end{eqnarray}

These three vector eigen function are perpendicular among themselves as well as with respect to each other [11]. In the form of matrices, vector Eigen functions can be written in this form,

\begin{eqnarray}
 L_{nhk}(\chi) =
\begin{pmatrix}
  \frac{\partial Z_{n})(\eta r)}{\partial r}P^{h}_{n}(\cos\theta)_{\sin}^{\cos} h \phi  \\
  \frac{Z_{n}(\eta r)}{r}P^{h}_{n}(\cos\theta)_{\sin}^{\cos} h \phi  \\
  \frac{h Z_{n}(\eta r)}{\sin\theta}P^{h}_{n}(\cos\theta)_{\cos}^{\sin} h \phi \\
 \end{pmatrix}
\end{eqnarray}

\begin{eqnarray}
 M_{nhk}(\chi) =
\begin{pmatrix}
   0  \\
  \mp\frac{h Z_{n}(\eta r)}{r}P^{h}_{n}(\cos\theta)_{\sin}^{\cos} h \phi  \\
  - Z_{n}(\eta r) \frac{\partial P^{h}_{n}(\cos\theta)_{\cos}^{\sin} h \phi} {\partial\theta} \\
 \end{pmatrix}
\end{eqnarray}

\begin{eqnarray}
 N_{nhk}(\chi) =
\begin{pmatrix}
  \frac{h Z_{n}(\eta r)}{k r}P^{h}_{n}(\cos\theta)_{\sin}^{\cos} h \phi  \\
  \frac{\partial_{n}(\eta r)}{kr \partial r}\frac{P^{h}_{n}(\cos\theta)_{\sin}^{\cos} h \phi} {\partial\theta}  \\
  \mp\frac{h}{\sin\theta}P^{h}_{n}(\cos\theta)_{\cos}^{\sin} h \phi \\
 \end{pmatrix}
\end{eqnarray}

\subsection{Scattering Superposition}

In scattering problems, it is desirable to determine an unknown scattered field  that is due to a known incident field. Using the principle of scattering superposition we can write Dyadic Green's equation as superposition of direct wave and scattering wave. In Figure 2, concept of scattering superposition is shown in which there is a sensor located inside the arm of body considered as sphere. The sensor transmits the wave to antenna which is divided in two parts as Direct wave and Scattered wave. The Direct wave is considered as wave directly transmits from sensor to transmitter and scattered wave is composed of reflection and transmission waves. Therefore, general equation of scattering superposition is illustrated as:

\begin{eqnarray}
G(x,x_0)= G_{d}(x,x_0) + G_{s}(x,x_0)
\end{eqnarray}

\begin{figure*}[!t]
\centering
\includegraphics[height=9cm, width=18cm]{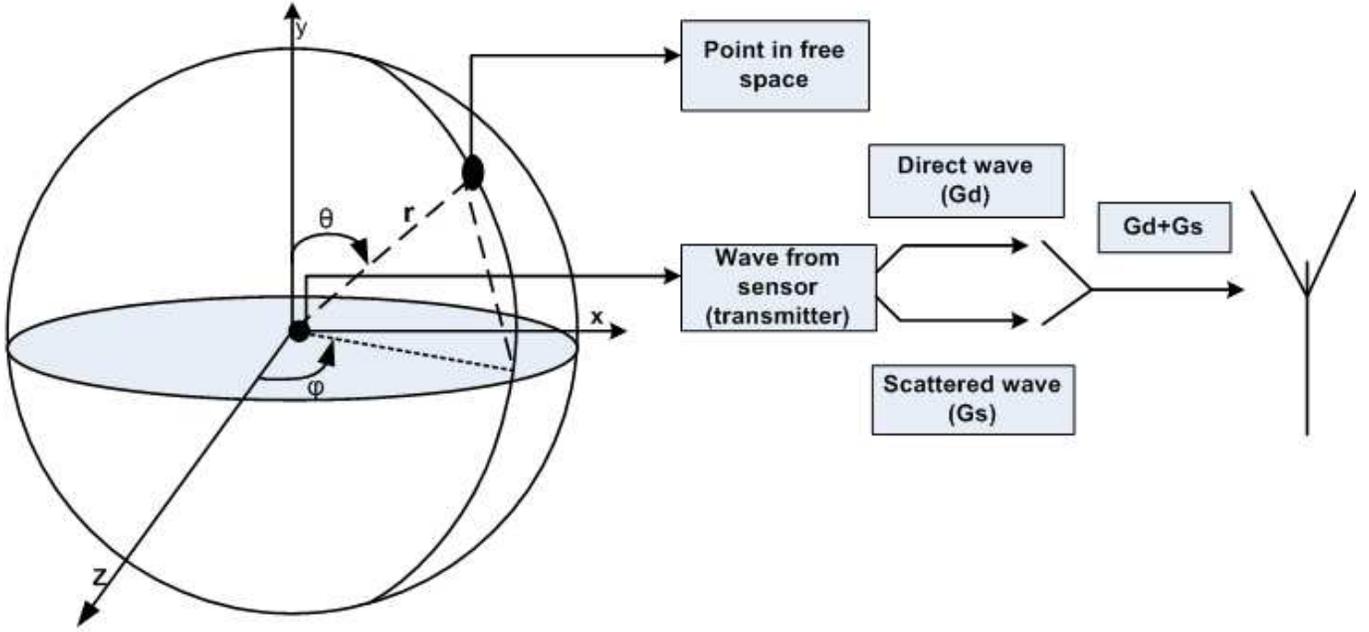}
\caption{ \textbf{Scattering Superposition} }
\end{figure*}

Dyadic Green's equation is divided in to two parts as direct wave $[G_d(x,x_0)]$ and scattered wave $[G_s(x,x_0)]$. The direct wave corresponds to direct from source to measuring point and scattered is the reflection and transmission waves due to presence of dielectric interface.

\subsection{Superposition of Direct Wave}

The direct component of DGF is given as [11]:

\begin{eqnarray}
\begin{split}
G_{d}(x,x_0)= \frac{rr}{k^2}(\delta(x-x_0)+\frac{\jmath}{8\pi}\int_{-\infty}^{\infty} dh\sum_{n=-\infty}^{\infty}\frac{1}{n^2} x  \times\\
\begin{cases}
    M^{(1)}_{nhk}(X)\bigotimes M_{nhk}^\ast(X_0)+ N^{(1)}_{nhk}(X)\bigotimes N_{nhk}^\ast(X_0)\\
    M_{nhk}(X)\bigotimes M^{(1\ast)}_{nhk}(X_0)+ N_{nhk}(X)\bigotimes N^{(1\ast)}_{nhk}(X_0)\\
\end{cases}
\end {split}
\end{eqnarray}

In the above equation of DGF, $r>r_0$ is for first case and $r<r_0$ is second case.The $\ast$ denotes the conjugation and $\bigotimes$ is for the Dyadic product.  Here we introduces superscript (1) for outgoings wave and other for standing waves. If the vector eigen function has the superscript (1) then, $H^{(1)}_{n}$ is chosen for $Z_{n}$ and $J_{n}$ should be used otherwise.

\subsection{Superposition of Scattered Wave}

Here we discuss four different scenarios for the scattering components of DGF along with boundary conditions $G_s {(x,x_0)}$. (i) Both receiver and transmitter are inside the body. (ii) The receiver is located outside and transmitter is located inside the body. (iii) The receiver is located inside and transmitter is outside the body. (iv) Both transmitter and receiver are located outside the body.
Receiver and transmitter are in the order: $1$ denotes the medium inside human body and $2$ is for free space medium.

\subsection{Transmitter and Receiver Located Inside Body}
In this case, Receiver and Transmitter both located inside the body so we can write Dyadic Green's equation as,

\begin{eqnarray}
 \begin{split}
G^{(11)}_{s}(x,x_0)=\frac{\jmath}{8\pi}\int_{-\infty}^{\infty} dh\sum_{-\infty}^{\infty}\frac{1}{\eta^2} x\\
 \times [M_{nhk1} N_{nhk1}] R_12 \times
\begin{cases}
    N_{nhk1}(X_0)^T\\
    M_{nhk1}(X_0)^T
\end{cases}
\end{split}
\end{eqnarray}

$R_{12}$ contains reflection coefficients. $R_{12}$ is calculated in literature using boundary conditions, its matrix is given by [16]:

\begin{eqnarray}
 \begin{split}
R_{12}= [J_n(\eta_1d)H_n(\eta_2d)-H_n(\eta_2d)J_n(\eta_1d)]^-1\\
\times[H_n(\eta_2d)H_n(\eta_1d)-H_n(\eta_1d)J_n(\eta_2d)]^-1
\end{split}
\end{eqnarray}

In the above equation of reflection coefficient 'd' represents radius of spherical body model, $\eta^{2}_{1}=k^{2}_{1}-h^2,\eta^{2}_{2}=k^{2}_{2}-h^2,k_{1}^{2}=\omega^{2}\mu_{1}\epsilon_{1},k_{2}^{2}=\omega^{2}\mu_{2}\epsilon_{2}$. The $2x2$ matrices for  $j_{n}(\eta d)$ and $H_{n}(\eta d)$ are expressed as:

\begin{eqnarray}
\begin{split}
 B_{n}(\eta_{p} d) = \frac{1}{\eta_{p}^{2} d}\times
 \begin{pmatrix}
  \jmath\omega\epsilon_{p}\eta_{p}d B_{n}(\eta_{p} d)& -nh  B_{n}(\eta_{p}  \\
  \ -nh B_{n}(\eta_{p} & - \jmath\omega\mu_{p}\eta_{p}d B_{n}(\eta_{p} d) \\
 \end{pmatrix}
 \end{split}
\end{eqnarray}
$B_{n}$ is either $H_{n}^{(1 or J_{n})}$, $B(.)$ is the derivative of $B$ w.r.t the whole argument, and p=1,2

\subsection{Transmitter Located Inside and Receiver Located Outside Body}

In this case DGF can be written as :

 \begin{eqnarray}
 \begin{split}
G_{s}^{(21)}(x,x_0)=\frac{\jmath}{8\Pi}\int_{-\infty}^{\infty} dh\sum_{n=-\infty}^{\infty}\frac{1}{\eta^2}\\
 \times [ N_{nhk}.  M_{nhk}]T_{12}
\begin{pmatrix}
   N^{\ast}_{nhk1}(x_0)^{T}\\
   M^{\ast}_{nhk1}(x_0)^{T} \\
  \end{pmatrix}
  \end{split}
\end{eqnarray}

In the above equation $T_{12}$ is a transmission coefficient Matrix and given as:

 \begin{eqnarray}
 \begin{split}
T12=\frac{2\omega}{\pi\eta_{1}^{2} d} [J_n(\eta_1 d )H_n(\eta_2 d)-H_n(\eta_2 d)J_n(\eta_1 d)]^-1\\
\times
\begin{pmatrix}
   \varepsilon_{1}&  0\\
   0&\varepsilon \\
  \end{pmatrix}
  \end{split}
\end{eqnarray}

\subsection{Both Transmitter and Receiver Located Outside Body}

 \begin{eqnarray}
 \begin{split}
G_{s}(x,x_0)=\frac{\jmath}{8\Pi}\int_{-\infty}^{\infty} dh\sum_{n=-\infty}^{\infty}\frac{1}{n^2}\\
 \times [M_{nhk} N_{nhk}] R_21\\
\begin{cases}
    N_nhk(X_0)^T
    M_nhk(X_0)^T
\end{cases}
\end{split}
\end{eqnarray}

Similarly as $R_{12}$, $R_{21}$ is the reflection coefficient matrix and it is given as:

 \begin{eqnarray}
 \begin{split}
R21= [J_n(\eta_1d)H_n(\eta_2d)-H_n(\eta_2d)J_n(\eta_1d)]^-1\\
\times[J_n(\eta_2d)J_n(\eta_1d)-J_n(\eta_1d)J_n(\eta_2d)]
\end{split}
\end{eqnarray}

%\begin{eqnarray}
%\begin{split}
%% \nonumber to remove numbering (before each equation)
%     \times
%\begin{cases}
%    N_nhk(X_0)^T
%    M_nhk(X_0)^T
%\end{cases}
%\end{split}
%\end{eqnarray}

\subsection{Transmitter Located Outside and Receiver Inside Body}
In this case, we can write DGF as:

\begin{eqnarray}
\begin{split}
G_{s}(x,x_0)=\frac{\jmath}{8\Pi}\int_{-\infty}^{\infty} dh\sum_{n=-\infty}^{\infty}\frac{1}{n^2}\\
 \times [M_{nhk} N_{nhk}] T_{21}\\
 \begin{pmatrix}
   N^{\ast}_{nhk1}(x_0)^{T}\\
   M^{\ast}_{nhk1}(x_0)^{T} \\
  \end{pmatrix}
\end{split}
\end{eqnarray}

$T_{12}$ is the transmission coefficient matrix, given as:

 \begin{eqnarray}
 \begin{split}
T21= \frac{2\omega}{\Pi \eta d}[J_n(\eta_1 d)H_n(\eta_2 d)-H_n(\eta_2 d)J_n(\eta_{2}^{2} d)]^-1\\
     \times
\begin{pmatrix}
   \varepsilon_{2}&  0\\
   0&-\mu_{2}\\
  \end{pmatrix}
\end{split}
\end{eqnarray}

\section{Transmitter and Receiver Located Outside of the Body}

In this section we presents the equation which is required for simulation. With the help of simulation it will be easy to study the propagation characteristics of arm motion making spherical pattern.

\begin{eqnarray}
\begin{split}
G_{s}(x,x_0)=\frac{\jmath}{8\Pi}\int_{-\infty}^{\infty} dh\\
\sum_{n=-\infty}^{\infty}\frac{1}{n^2} G_{nh}(x,x_0) dh\\
\end{split}
\end{eqnarray}

$G_{nh}(x,x_0)$ is stated as:

 \begin{eqnarray}
 \begin{split}
 G_{nh}(x,x_0)=
\begin{pmatrix}
    N_{nhk}(X)^1 &  M_{nhk}(X)^1
\end{pmatrix}
\times R21 \\
\begin{pmatrix}
    N_{nhk}(X_0)^T
    M_{nhk}(X_0)^T
\end{pmatrix}
\end{split}
\end{eqnarray}

\section{Simulations}

As we have defined earlier, arm motion at different angles are presenting spherical pattern. Therefore, we simulate the radio propagation environment having radius  $ d=15cm $, megnatic permeability for human body (assume that permeability of human body is approximately equal to air) $\mu_{2}=1.256 \times 10^-6$,  similarly electric permittivity $\varepsilon_{2}=2.563\times10^-10$. The dielectric constant is mean value of all tissues of human body.
We take the surrounding homogeneous medium to be air with megnatic permeability $\mu_{1}=1.256 \times 10^-6$ and electric permittivity $\varepsilon_{1}=8.8542\times10^-12$. Frequency up to GHz is used for BAN communication, which is for ISM band. The Transmission frequency for simulation is 1GHz. We assumed that the transmitter is acting as point source at $x_{0}=(16cm,\frac{\pi}{2},0)$. The radial distance of receiver is $ r_{0}=18cm$ from the central spherical axis of shoulder. For the simulation, we assumed that receiver move along the azimuthal angle for varying values of $\phi_{0}$ and different heights from the center of shoulder.

For simulation, we consider equation (25) in which $G_{nh}(x,x_0)$ is used in matrix form of eigen functions. This equation has an integration which is not possible so we approximate it to summation. Thus, we approximate equation (25) in to this form:

\begin{eqnarray}
\begin{split}
G_{s}(x,x_0)=\frac{\jmath}{8\Pi}\sum_{l=-L}^{L}\\
\sum_{n=-Q}^{Q}\frac{1}{n^2} G_{nh}(x,x_0) dh \\
\end{split}
\end{eqnarray}

$L$ and $Q$ are the truncation limits and $\Delta H$ are the step size of integration. $N$ and $\Delta H$  are so small that could be ignored and has no effect on calculations. We only presents electric propagation of multi-path reflection and transmission waves of scattering DGF.This is more significant to represent the attribute of arm motion as compared to the direct DGF. Figure 2,3 and 4 show the scattering DGF (simulation) of electric field with the change in $\theta$.

\begin{figure}[h!]
%Require\usepackage{graphicx}
\includegraphics[height=9cm, width=9cm]{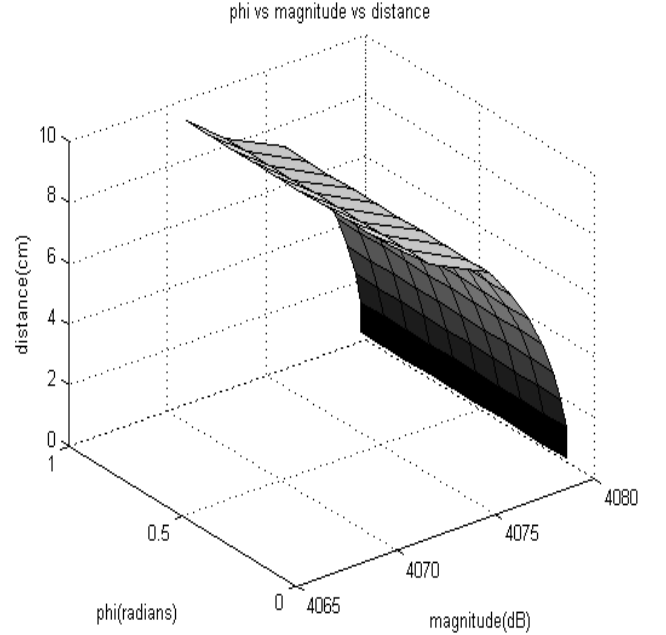}
\caption{ Magnitude of scattered field component $E_{\phi}$ versus angle $\phi$,with different values of $d$ and the angle is $\theta=\frac{\pi}{6}$ }
\end{figure}

Using equation (27), we have three components in $r$,$\theta$ and $\phi$ direction. Every Component of electric field is plotted as a function of azimuthal angle $\phi$. The values of $\phi$ is (0 to 2$\pi$), whereas at z coordinate different values of receiver has been plotted. The electric field is plotted, which is vector addition of three components. These all parameters are shown in the simulation graph.

By taking the value of $\theta=\frac{\pi}{6}$, figure $2$ shows that magnitude of electric field $(E_{\phi})$ is decreasing as the distance of receiving antenna is increasing from the sensor (transmitting antenna). The plot shows electric field component at different values of $\phi$, varying from $0$ to $2 \pi$. In this case, $E_{\phi}$ is decreasing from  ($4080$ to $4065$)dB by replacing the receiving antenna from $0$ cm to $10$ cm.

\begin{figure}[h!]
%Require\usepackage{graphicx}
\includegraphics[height=9cm, width=9cm]{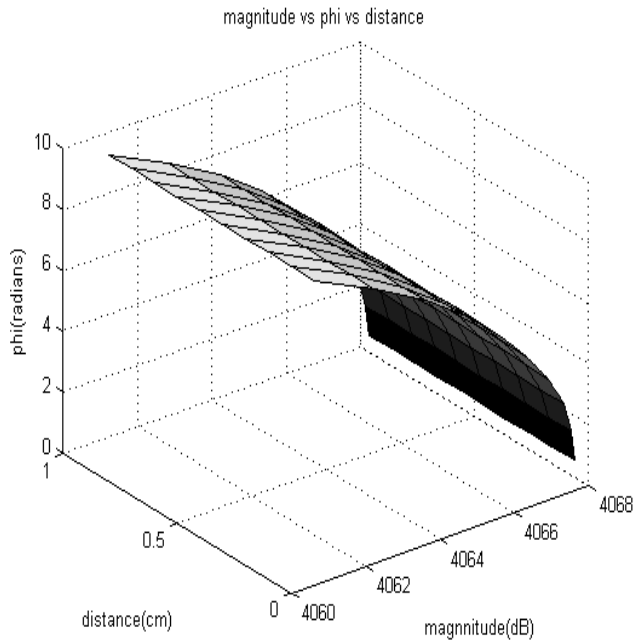}
\caption{Magnitude of scattered field component $E_{\phi}$ versus angle $\phi$,with different values of $d$ and the angle is $\theta=\frac{\pi}{3}$ }
\end{figure}

In Figure $3$, when we take value of $\theta=\frac{\pi}{3}$, magnitude of electric field $(E_{\phi})$ again decreases as the antenna moves away from sensor. For the values of $\phi$ from $0$ to $2 \pi$, $E_{\phi}$ has different values from ($4060$ from $4068$)dB. By changing position of receiving antenna from $0$ cm to $10$ cm.

\begin{figure}[h!]
%Require\usepackage{graphicx}
\includegraphics[height=9cm, width=9cm]{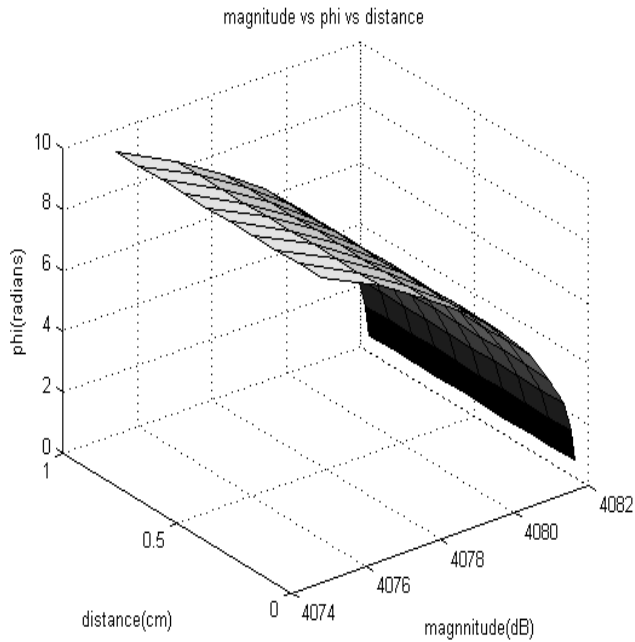}
\caption{Magnitude of scattered field component $E_{\phi}$ versus angle $\phi$,with different values of $d$ and the angle is $\theta=\pi $  }
\end{figure}

The values of distance  and $\phi$ are same, as described in the above graphs by only replacing the parameter $\theta=\pi$. Similarly in figure $4$ values of $E_{\phi}$ change from ($4082$ to $4074$)dB by moving the position of receiver away from transmitting antenna, which in return decreases the electric field intensity.

\section{Conclusion}

We have proposed a generic approach to derive an analytical channel modeling and propagation characteristics of arm motion as spherical model. To predict the electric field around body, we have formulated a two step procedure based on Dyadic Green's function. First, we derive Eigen functions of spherical model then calculated the scattering superposition to come across reflection and transmission waves of antenna. The model includes four cases where transmitter or receiver is located inside or outside of the body. This model is presented to understand complex problem of wave propagation in and around arm of human body. Simulation shows that Electric field decreases when receiver moves away from the shoulder with change of angle $\theta$.


\begin{thebibliography}{1}

\bibitem{1} T. Zasowski, F. Althaus, M. Stager, A. Wittneben, and G. Troster, "Uwb
     for noninvasive wireless body area networks: Channel measurements and
     results," Proc. IEEE Conf. on Ultra Wideband Systems and Technologies,
     pp. 285-289, Nov 2003.

\bibitem{2} A. Fort, J. Ryckaert, C. Desset, P.D. Doncker, P. Wambacq, and L.V.
     Biesen, "Ultra-wideband channel model for communication around the
     human body," IEEE Journal on Selected Areas in Communications, vol.
     24, no. 4, pp. 927-933, April 2006.


\bibitem{3} H. Ghannoum, C. Roblin, and X. Begaud,"Inves-
     tigationoftheuwbon-bodypropagationchannel,"
     http://uei.ensta.fr/roblin/papers/WPMC2006HGBANmodel.pdf, 2005.


\bibitem{4} D. Nierynck, C. Williams, and A. Nix, M. Beach,
     "Channelcharacterisationforpersonalareanetworks,"
     http://rose.bris.ac.uk/dspace/bitstream/1983/893/1/TD-05-115.pdf,
     Nov. 2007.


\bibitem{5} A. Alomainy, Y. Hao, X. Hu, C.G. Parini, and P.S. Hall, "Uwb on-
     body radio propagation and system modelling for wireless body-centric
     networks," IEE Proc. Commun., vol. 153, no. 1, pp. 107-114, 2006.

\bibitem{6} Y. Zhao, Y. Hao, A. Alomainy, and C. Parini, "Uwb on-body radio
     channel modelling using ray theory and sub-band fdtd method," IEEE
     Trans. On Microwave Theory and Techniques, Special Issue on Ultra-
     Wideband, vol. 54, no. 4, pp. 1827-1835, 2006.

\bibitem{7} J. Ryckaert, P.D. Doncker, R. Meys, A.D.L. Hoye, and S. Donnay,
     "Channel model for wireless communication around human body,"
     Electronic Letters, vol. 40, no. 9, 2004.

\bibitem{8} I.Z. Kovacs, G.F. Pedersen, P.C.F. Eggers, and K. Olesen, "Ultra
     wideband radio propagation in body area network scenarios," IEEE
     8th Intl. symp. on Spread Spectrum Techniques and Applications, pp.
     102-106, 2004.

\bibitem{9} J.A. Ruiz, J. Xu, and S. Shiamamoto, "Propagation characteristics of
     intra-body communications for body area networks," 3rd IEEE Conf. on
     Consumer Communications and Networking, vol. 1, pp. 509-503, 2006.

\bibitem{10} T. Zasowski, G. Meyer, F. Althaus, and A. Wittneben, "Propagation
       effects in uwb body area networks," IEEE Intrenational Conference on
       7UWB, pp. 16-21, 2005.

\bibitem{11} Z. Xiang and Y. Lu, "Electromagnetic dyadic green's function in
       cylindrically multilayered media," IEEE Trans. on Microwave Theory
       and Techniques, vol. 44, no. 4, pp. 614-621, 1996.

\bibitem{12} P.G. Cottis, G.E. Chatzarakis, and N.K. Uzunoglu, "Electromagnetic
       energy deposition inside a three-layer cylindrical human body model
       caused by near-?eld radiators," IEEE Trans. on Microwave Theory and
       Techniques, vol. 38, no. 8, pp. 415-436, 1990.

\bibitem{13} S.M.S Reyhani and R.J. Glover, "Electromagnetic modeling of spherical
       head using dyadic green's function," IEE Journal, , no. 1999/043, pp.
       8/1-8/5, 1999.

\bibitem{14} T.Zasowski, F. Althaus, M. Stager, A. Wittneben and G. Troster, "UWB
     for noninvasive wireless body area networks: channel measurement and
     results."in 2003 IEEE  conference on Ultra-Wide band  system and
     technologies,2003.pp.285-289.

\bibitem{15} A. Alomainy, Y. Hao, X.Hu,C.G. Parini and P.S. Hall, "UWB on-body
     radio propagation and system modeling for body centric networks,"
     in IEEE communication proceeding, vol. 153, no. 1, February 2006,
     pp. 107-114.
\bibitem{16} D. A. Macnamara, C, Pistorius and J. Malherbe, In troduction to
     the uniform geometrical theory of diffraction. Artech House:Boston, 1991.

\bibitem{17} C.T. Tai, Dyadic Green's Functions in Electromagnetic Theory, IEEE,
       New York, 1993.

\bibitem{18} Le-Wei Li, Senior Member, IEEE, Mook-Seng Leong, Senior Member, IEEE,
     Pang-Shyan Kooi, Member, IEEE,and Tat-Soon Yeo, Senior Member, IEEE

\bibitem{19} Astha Gupta, Thushara D. Abhayapala, " Body Area Networks:
     Radio Channel Modelling and Propagation Charaterstics".

\end{thebibliography}
\end{document}